# Probing microwave capacitance of self-assembled quantum dots

Guanglei Cheng[1], Jeremy Levy[1] and Gilberto Medeiros-Ribeiro[2,3]
[1]*Department of Physics and Astronomy, University of Pittsburgh, PA 15260*
[2]*Laboratorio Nacional de Luz Sincrotron, Campinas, SP 13083-100, Brazil*
[3] *Hewlett-Packard Labs, Palo Alto, CA 94304*

Self-assembled quantum dots have remarkable optical and electronic properties that make them leading candidates for spin-based quantum information technologies. The characterization of such systems requires rapid and local determination of both charge and spin degrees of freedom. We present a way to probe the capacitance of small ensembles of quantum dots at microwave frequencies, which can provide important information about electron occupation and vertical tunneling of electrons or holes in the quantum dots. The technique employs a capacitance sensor based on a microwave microstrip resonator with sensitivity $\sim 10^{-19}$ F/$\sqrt{\mathrm{Hz}}$, high enough to probe single electrons. The integration of this design in a scanning microscope will provide an important tool for investigating single charge and spin dynamics in self-assembled quantum dot systems.

Self-assembled quantum dots (SAQDs) have achieved extensive interests on potential applications in spin-based quantum information technologies due to their atomic-like quantized energy levels, scalability, and relatively long coherence times. Single spin detection and manipulation in SAQDs have been successfully performed optically using Faraday and Kerr rotation techniques[1-4]. In those works, single spins are addressed by a laser spot of ~1 μm size and manipulated by picosecond laser pulses in a very low density SAQDs sample. An all-electrical means of information processing that does not require optical preparation, manipulation and interrogation methods may render a particular device more functionality and better perspectives insofar as integration is concerned. One way to electrically address and control single spins is to use a nanometer-scale metallic probe as a top gate and measure the quantum capacitance as a figure of merit for controlling the charge states of electrons in SAQDs.

The capacitance of a single SAQD is normally on the order of $10^{-18}$ F. Within this regime, conventional capacitance sensors like L-C circuits and capacitance bridges are not sensitive enough. A configuration with leads to a device containing very few SAQDs would also exhibit a parasitic capacitance orders of magnitude larger than the SAQDs capacitance themselves. Another approach to capacitance sensing is based on a microstrip resonator design that was first developed in a commercial product, RCA's CED VideoDisc[5, 6]. The resonant frequency of the microstrip resonator is extremely sensitive to a capacitance change, if a capacitor is attached. An increase in the contact capacitance will lower the resonant frequency accordingly. And this frequency shift will be detected by driving the circuit off resonance and measuring the resulting change in transmitted microwave power using a schottky peak detector. The sensitivity has been reported from RCA's $10^{-19}\ \mathrm{F}/\sqrt{\mathrm{Hz}}$ to "sub-zeptofarad" ($10^{-21}\ \mathrm{F}/\sqrt{\mathrm{Hz}}$)[7, 8].

Here we investigate the capacitance-voltage characteristics of Metal-Insulator-Semiconductor devices containing SAQDs by a microwave capacitance sensor mentioned above. The samples are grown by molecular beam epitaxy (MBE), consisting of an undoped 1000 nm-thick GaAs buffer layer grown at 600 °C under an As flux of $10^{-6}$ Torr, followed by an 80 nm Si-doped back contact layer with a nominal concentration of $\sim 1\mathrm{x}10^{18}$ $\mathrm{cm}^{-3}$. The temperature is then lowered to 530 °C, and an undoped GaAs tunneling barrier of thickness $t_b$=16 nm separates the InAs SAQDs from the back contact. The InAs QDs are subsequently grown by pulsing the In flux at an effective growth rate of 0.01 ML/s until island nucleation. The typical QD density is about $1\mathrm{x}10^{10}$ $\mathrm{cm}^{-2}$. Then a 6 nm thick strain-reducing layer (SRL) of $In_{0.2}Ga_{0.8}As$ is deposited over the SAQDs. At last, an undoped GaAs cap layer is grown and the temperature is ramped up to 600 °C[9]. Schottky and ohmic contacts with an area of $200 \times 200\ \mu\mathrm{m}^2$ are also defined by conventional optical lithography processes.

A schematic of the experimental setup for the microwave measurements is shown in Fig. 1a. The microwave measurements are performed using a home-designed microstrip resonator, with a design similar to that of Ref. 8. The resonant frequency is about $f_0$=1.85 GHz unloaded (i.e., no sample attached). Connecting the sample directly causes too large of a resonance frequency shift, so a $C_0$=0.1 pF series capacitor is used to desensitize the resonator. The resonant frequency shifts by about 250 MHz when this capacitor is attached. Operating frequencies between 1.4 GHz to 1.8 GHz are optimal for achieving high measurement sensitivity. A –20 dbm RF signal with fixed frequency is fed from one side of the resonator and the transmitted power is measured by a broadband schottky peak detector (Agilent 8473B) in the other side. A low input RF power is preferred to prevent sample heating. The top gate of the SAQDs structure is connected in series to the 0.1 pF capacitor, which is in turn connected to the central resonator microstrip. The capacitance change $\Delta C$ in SAQDs caused by quasi-static DC bias will be also reduced by a factor of $(C_0/C_q)^2$, where $C_q$ is the capacitance of quantum dots. Assuming a typical $C_q$=10 pF of this sample device, the capacitance change seen by the resonator will be $1/10^4$ of the original quantum capacitance variation. A DC bias voltage with a small AC modulation (10 KHz, 0.05 $V_{rms}$) is applied through a 100 nH inductor, which can let low frequency AC modulation pass through while still blocking the RF transmission. The output of the peak detector is lock-in detected at 10 KHz modulation frequency. This type of modulation gives a low frequency perturbation on microwave capacitance and the resulting differential $dC/dV_b$ curve can be integrated to yield the microwave C-V response.

A series of resonators are fabricated, and the sensitivity of a typical resonator is calibrated. In this calibration, the SAQDs sample is replaced by a commercial varactor (On Semiconductor, MMBV3102LT1G) to allow precise capacitance tuning and calibration. By tuning the varactor from 4 pF to 30 pF, which corresponds an overall series capacitance ranging from 97.56 fF to 99.67 fF, the resonant frequency shifts from 1573.5 MHz to 1566.6 MHz, as shown in Fig. 1b. The frequency sensitivity $df/dC$ in this operating region is then determined as 3.3 MHz/fF through these results. The voltage sensitivity $dV/df$, which is a figure of merit of the peak detector, gives the voltage response due to a change in frequency and is determined by measuring the slope of the resonance curve at the working frequency. For fixed input power -20 dbm, the voltage sensitivity is about 12.5 μV/MHz at -3 db points on the slopes of the resonance curves. Given an overall measured system noise level $V_{noise} = 25\ \mathrm{nV}/\sqrt{\mathrm{Hz}}$ at 10 KHz, the system sensitivity is calculated to be $\delta C = \dfrac{V_{noise}}{df/dC \bullet dV/df} = 6.1\times 10^{-19}\ \mathrm{F}/\sqrt{\mathrm{Hz}}$

As a reference, we first investigate the conventional low-frequency capacitance-voltage (C-V) spectrum by sweeping the DC voltage $V_b$ with a small AC modulation $V_{ac}$ of frequency $f_m$, and measuring the in-phase and out-of-phase responses with a dual-channel lock-in amplifier[10, 11]. Fig. 2 shows a C-V curve obtained at $f_m$=5 kHz. The two s-shell states are readily resolved at DC bias voltages $V_{s1}$= –1.365 V and $V_{s2}$= –1.165 V respectively. The four p-shell states are located between –0.9 V and –0.5 V, and are not individually resolvable owing to peak broadening caused by quantum dot size dispersion and the large number of dots probed[12].

The microwave capacitance of SAQDs sample with 16 nm thin tunneling barrier is then measured at $T$=10 K in a He flow cryostat. Using similar voltage bias scheme as the low frequency measurement, DC bias $V_b$ plus 0.05 $V_{rms}$ AC voltage modulation is applied via the RF choke. To determine the working microwave frequency, $V_b$ is set to –0.3 V, where a maximum slope $dC/dV_b$ is expected according to the low frequency C-V curve (Fig. 2). Under this condition, the microwave frequency is then swept to obtain a differential resonance curve $dV/df$ with $f$ the microwave frequency, as shown in the inset of Fig. 3a. Two types of information can be extracted from this curve. First it provides a signature of sensitivity of the resonator. Failure in seeing this curve means the resonator cannot sense the frequency shifts caused by the small AC voltage modulation. In the mean time as a second order effect, it can also represent a loading of the cavity, due to a leaky sample or a sample with very large capacitance. That is why the $C_0$ capacitor is so important to fine tuning the set up. Second the microwave working frequency can be set at the two extreme values of this curve, which indicate optimal sensitivities. Once the working frequency is determined, the bias voltage $V_b$ is then swept, resulting in a differential $dC/dV_b$ curve (Fig. 3a), which can be integrated to obtain the microwave C-V spectrum, as shown in Fig. 3b. After a quick comparison between Fig. 2 and Fig. 3b, a conclusion can be drawn that the C-V spectra are almost frequency independent from KHz to GHz regime for this SAQDs sample, which contrasts previous results with thicker tunneling barriers[11]. In a similar measurement for SAQDs sample with a thicker barrier of 40 nm, we can't even retrieve the differential resonance curve, which indicates that the tunneling barrier thickness strongly influences the microwave response of the electrons since the tunneling time is exponentially dependent on tunneling barrier thickness.

In summary, we utilize a microstrip resonator based capacitance sensor to measure SAQDs capacitance at microwave frequencies. By comparing with low frequency C-V curve, we see the spectra are frequency-independent owing to the very thin tunneling barrier of this SAQDs sample. The sensitivity of the microstrip capacitance sensor is on the order of $10^{-19}$ F/$\sqrt{\text{Hz}}$, which can be further improved by using the latest microwave power detectors, and is enough to probe the capacitance of a single quantum dot (~$10^{-18}$ F). The sensitivity of this technique is also "tunable" by adjusting the series capacitance $C_0$, provided that the overall capacitance of the sample is not too large. This technique can be used in conjunction with a low temperature scanning force microscope[13] to obtain high-resolution images of the microwave response, and can also be used to gate SAQD structures at these frequencies, with the ultimate goal of electrically sensing and controlling a single SAQD.

This work was supported by NSF Materials World Network grant DMR-0602846 and CNPq (Brazil) CIAM grant.

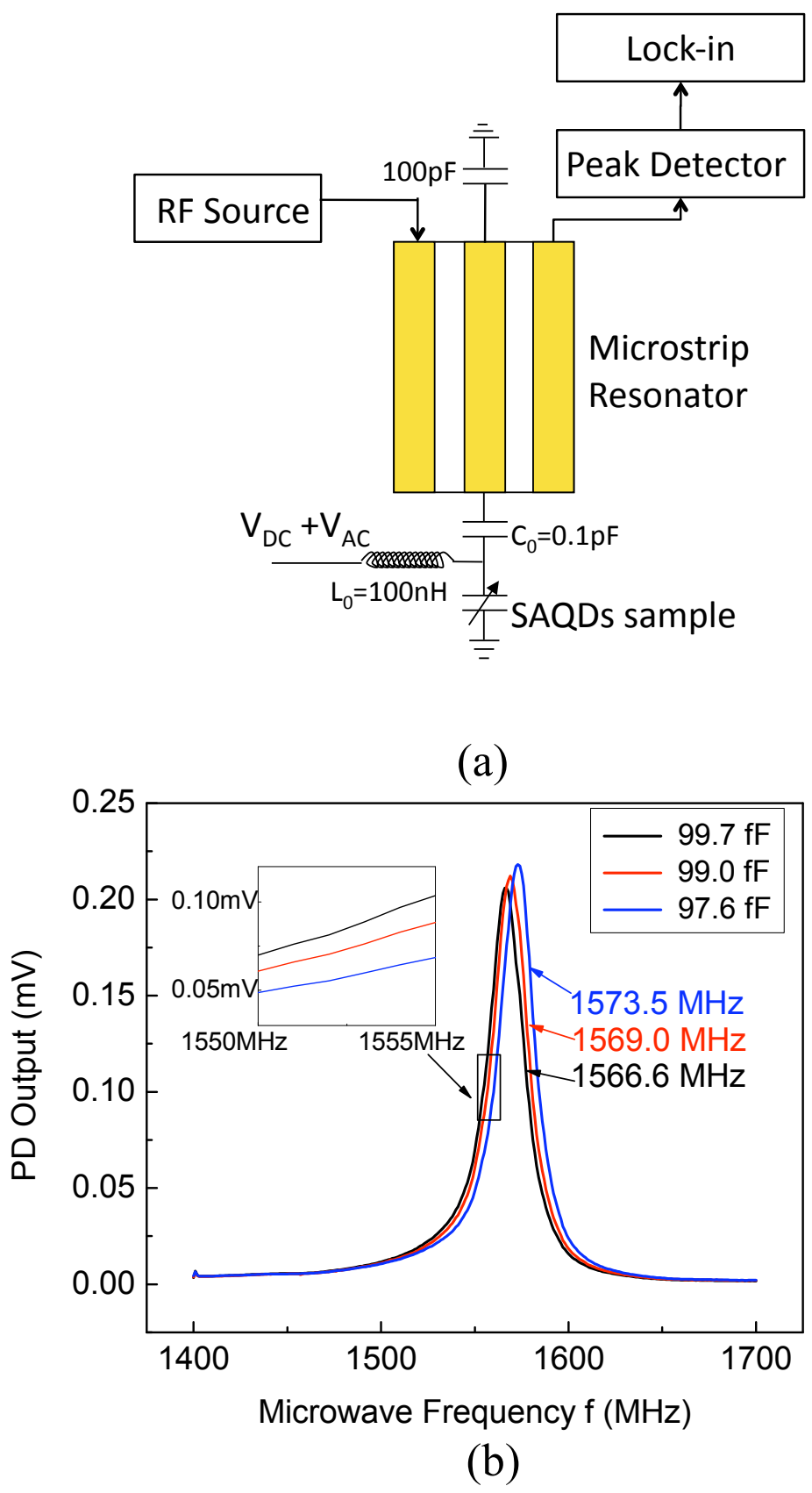

Fig. 1. (Color online) (a) Experimental setup. The resonance of this microstrip resonator is shifted by varying DC bias applied to the SAQDs sample. This frequency shift is then sensed by the schottky peak detector, by converting microwave power to voltage. (b) Resonance curves at different contact capacitance. The SAQDs sample is replaced by a varactor to allow determined capacitance tuning. The inset is a zoom-in image from 1550 MHz to 1555 MHz, showing frequency shifts will cause detectable voltage change at fixed microwave frequencies. The frequency sensitivity is determined to be 3.3 MHz/fF in this region.

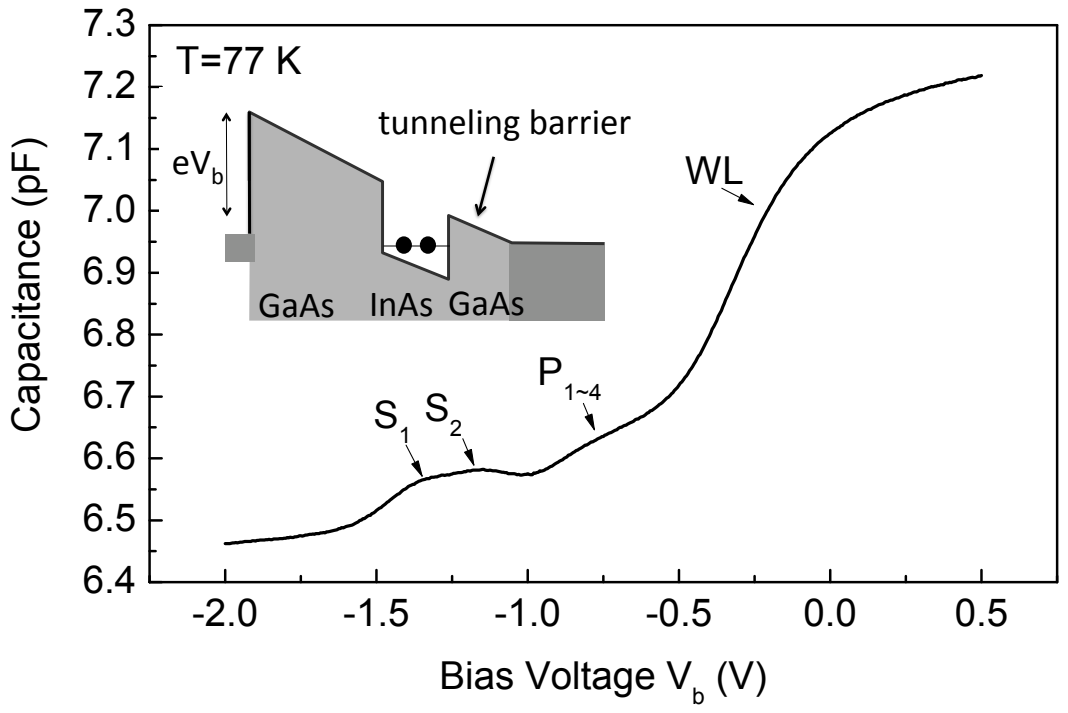

Fig. 2. Low frequency C-V curve. Two S states $S_1$ and $S_2$ are located at –1.365 V and –1.165 V. P states are degenerated due to SAQDs size dispersion and $10^6$ number of dots probed. The inset shows the band diagram of the SAQDs.

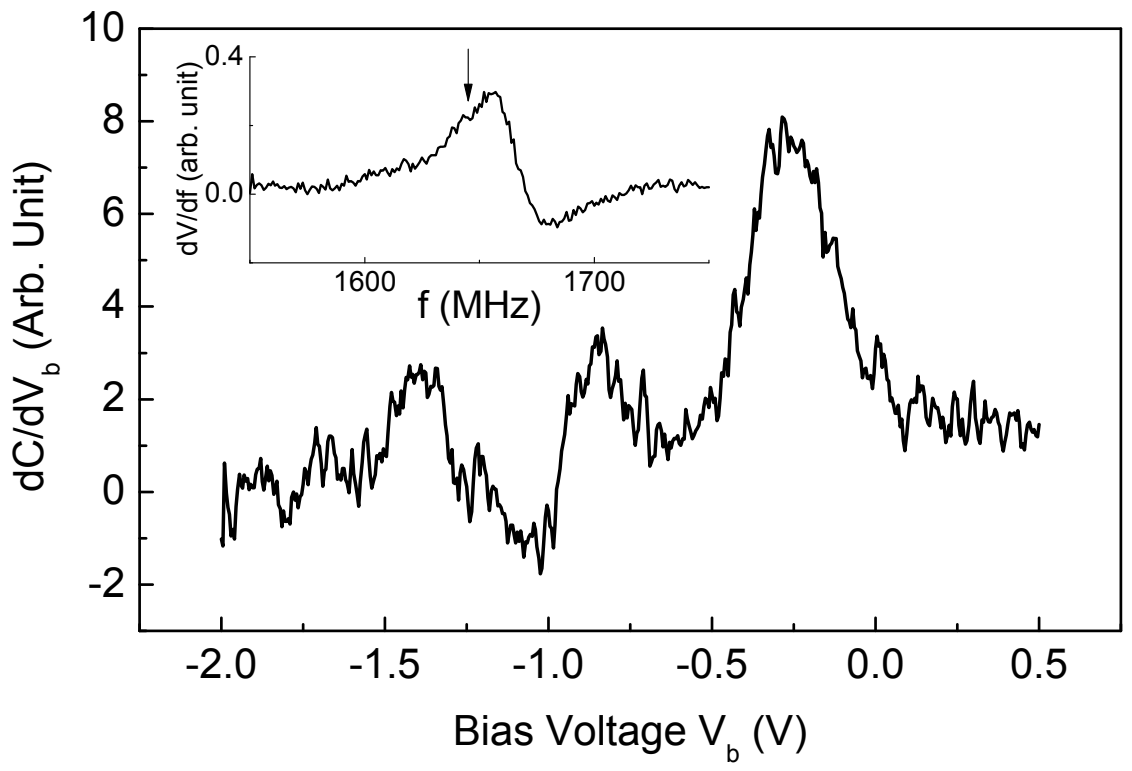

(a)

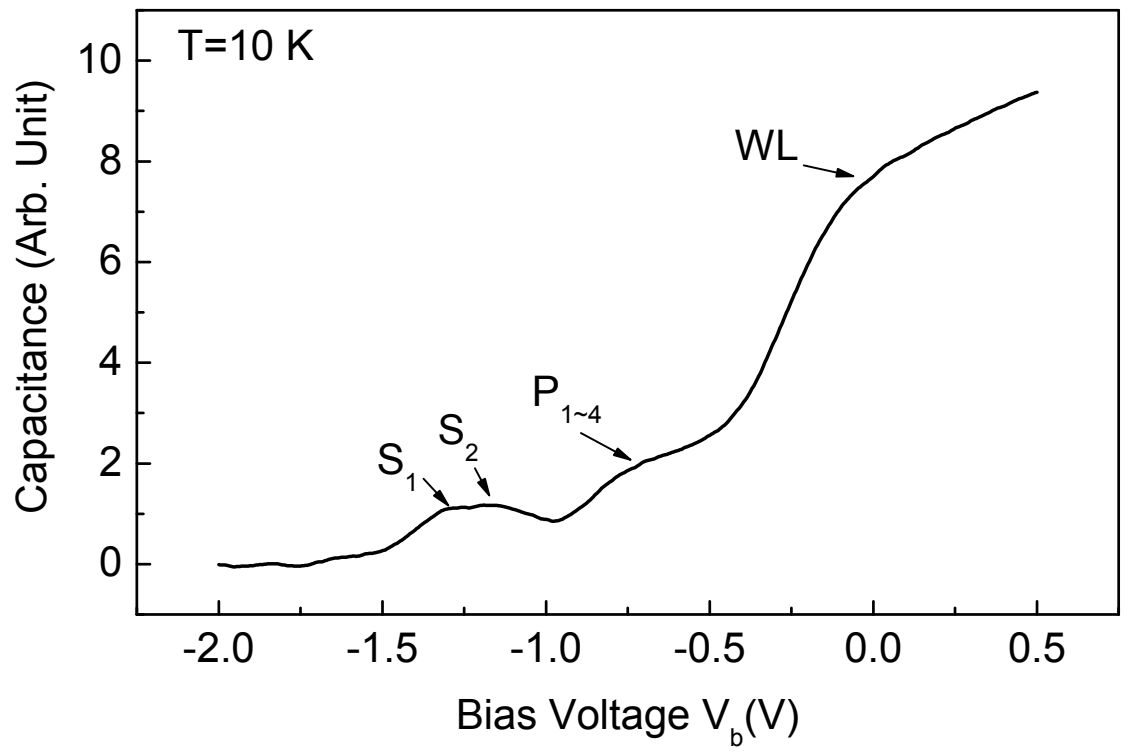

(b)

Fig.3. (a) Differential $dC/dV_b$ curve obtained by lock-in detecting the AC modulation frequency. This curve shows the slope of conventional C-V curve. The inset shows the differential resonance curve of the resonator. The arrow indicates the working frequency 1650 MHz chosen in this experiment. (b) C-V curve recovered by integrating the $dC/dV_b$ curve in (a).